# Performance Analysis of UWB Based Wireless Sensor Networks in Indoor Office LOS Environment

Wu Xuanli[1,2], Cao Yang[2], Yang Xiaozong[1], Luo Chao[2]
[1] School of Computer Science and Technology
[2] School of Electronics and Information Engineering
Harbin Institute of Technology
Harbin, China
xlwu2002@hit.edu.cn

*Abstract*—With the fast development of wireless sensor networks (WSN), more attentions are paid to high data rate transmission of WSN, and hence, in IEEE 802.15.4a standard, ultra-wideband (UWB) is introduced as one of the physical layer technique to support high transmission data rate and precisie locationing applications. In order to analyze the bit error rate (BER) performance of UWB based WSN, a system model considering intra-symbol interference (IASI), inter-symbol interference (ISI), multiuser interference (MUI) and addictive white Gaussian noise (AWGN) is proposed in this paper, and then verified using simulation results. Moreover, the pulse waveforms complying with the spectrum requirement of IEEE 802.15.4a standard are given, and based on such obtained pulses, the effect of transmission data rate and user number is also shown. Results show that with the increase of SNR, the intra-symbol interference will decrease the system performance significantly, and system performance can be improve by using pulse waveforms with little intra-symbol interference.

## I. INTRODUCTION

In recent years, with the fast development of microelectronics, computer science and wireless communication techniques, wireless sensor networks has attracted more attentation by connectting information world and physics world together, and then changed the way of interactions between human and nature [1]. Many literatures shows that ultrawideband (UWB) technique is one of the enabling technologies for WSN due to its advantages on potentially high transmission data rate, low power consumption and complexity, precise locationing and tracking ability, and little interference to other systems [2, 3]. Thanks to the above advantages of UWB technique, in 2007, UWB is selected as one of the physical layer technique for IEEE 802.15.4a standard [4] for high data rate applications. The performance of UWB in dense multipath enviroment has been investigated by many literatures. Reference [5] shows the performance of UWB based WSN for different application environments. Reference [6] analyzed the system performance of IR-UWB D-Rake receives over IEEE 802.15.4a multipath fading channels with norrow-band interference. Analysis of signal to interference and noise ratio (SINR) is givne in [7] with direct-sequence UWB system in generalized S-V channel. In reference [8], it is figured out that inter-symbol interference (ISI) can be neglected with low transmission data rate, while for high transmission data rate, ISI will influence system performance significantly.

However, the intra-symbol interference (IASI) is not considered in above literatures, and the bit error rate (BER) formulation considering MUI, ISI, IASI and additative white Gaussian noise (AWGN) has not been given as well. Hence, in this paper, the BER formulation considering MUI, ISI, IASI and AWGN simultaneity is first deduced in IEEE 802.15.4a indoor office environment, and then, the effect of system parameters like transmission data rate and pulse waveforms occuping different bandwidth are analyzed, and results shows that in indoor office enviroment, the intra-symbol interference decreased the system performance significantly enen in low data rate and small use number.

## II. CHANNLE MODEL OF IEEE 802.15.4A

In IEEE 802.15.4a standard, there are many types of channel models for UWB communication system, and in this paper, the indoor office LOS environment with path frequency dependence is used. Then, in time domain, the impulse response of UWB system can be written as follows:

$$h(t) = f(t) * \sum_{l=1}^{L}\sum_{k=1}^{K} \alpha_{k,l} \delta(t - T_l - \tau_{k,l}). \quad (1)$$

where, $L$ and $K$ denote the total number of clusters and rays, $\alpha_{k,l}$ is the cofficiency of the $k$-th ray in the $l$-th cluster, and $\tau_{k,l}$ is the arrival time of $k$-th ray relative to $l$-th cluster arrival time $T_l$.

In frequency domain, (1) can be rewritten as:

$$H(\omega) = F(\omega) \sum_{l=1}^{L}\sum_{k=1}^{K} \alpha_{k,l} \exp(-j\omega \cdot (T_l + \tau_{k,l})). \quad (2)$$

where, $F(\omega)$ denotes the frequency dependence of ray arrivals, which can be given by:

$$F(\omega) = C_0 (\omega / \omega_0)^{-\kappa}. \quad (3)$$

where, $C_0$ is a constant, $\kappa$ is the frequency dependence of the pathloss, and $\omega_0$ is the reference frequency. Moreover, using Taylor series, $F(\omega)$ can be expanded as:

$$F(\omega) \approx F(\omega_c) + F'(\omega_c)(\omega - \omega_c). \quad (4)$$

where, $\omega_c$ represents the center frequency.

In indoor office LOS environment, $\kappa$ is considerably small and $F(w)$ is a slowly varying function of $\kappa$ within the applied frequency bandwidth [8]. Thus we can obtain the approximation of $F(\omega) \approx F(\omega_0)$ by ignoring the higher order terms of Taylor series.

It is noted that for IR-UWB systems using pulse based transmitter and receiver, the pulse has only positive and negative polarity. Thus, there is no need to consider random phase angle in equation (1).

Without loss of generality, we suppose the first ray in the first cluster is the desired ray to be received, whose energy is $\Omega_0$. Then the delay of the $n$-th user from different propagation distance and different transmitting time can be expressed as: $t_u^{(n)} = \tau^{(n)} - \tau^{(1)}$. Moreover, for the same user the delay of the $l$-th cluster relative to the first cluster is $t_c^{(n)} = T_l - T_1$. And in the same cluster, the delay of the $k$-th ray relative to the first ray is $t_p^{(k)} = \tau_{k,l} - \tau_{1,l}$. $\tau_{code}^{(n)} = (C_{i,j}^{(n)} - C_{0,j}^{(n)}) \cdot T_c$ denotes the TH code interval between the $i$-th interfering pulse and the current receiving pulse with $T_c$ corresponding to the duration of time hopping code.

According to IEEE 802.15.4a indoor office channel model, the distribution of the cluster arrival times is given by a Poisson processes:

$$p(T_l / T_{l-1}) = \Lambda_l \exp[-\Lambda_l (T_l - T_{l-1})], l > 0. \quad (5)$$

where, $\Lambda_l$ is the cluster rate.

Similarly, the distribution of $\tau_{k,l}$ is modelled with a mixture of two Poisson processes as follows:

$$p(\tau_{k,l} / \tau_{(k-1),l}) = \beta \lambda_1 \exp[-\lambda_1 (\tau_{k,l} - \tau_{(k-1),l})] \\ + (1-\beta)\lambda_2 \exp[-\lambda_2 (\tau_{k,l} - \tau_{(k-1),l})], k > 0. \quad (6)$$

where, $\beta$ is the mixture probability, while $\lambda_1$ and $\lambda_2$ are the ray arrival rates.

According to probability theory, $t_c^{(l)}$ obeys Poisson distribution with two parameters $\Lambda$ and $l$, and the probability dense function (PDF) can be given by [8]

$$f_c(x) = \Lambda \exp(-\Lambda x) \frac{(\Lambda x)^{(l-2)}}{(l-2)!}. \quad (7)$$

Nevertheless, $t_p^{(k)}$ is described with a mixture Poisson processes and it is difficult to analyze its distribution function and probability dense function (PDF). Meanwhile, we notice that in indoor office LOS environment, $\beta$ is considerably small, which indicates that the occurrence of Poisson process with parameter $\lambda_1$ is very small and the Poisson process with parameter $\lambda_2$ is dominant. To simplify our computation, we take $t_p^{(k)}$ as a single Poisson process with parameter $\lambda_2$ and the PDF is then given by:

$$f_p(x) = \lambda_2 \exp(-\lambda_2 x) \frac{(\lambda_2 x)^{(k-2)}}{(k-2)!}. \quad (8)$$

The PDF of $\tau_{code(i)}^{(n)}$ can be expressed as:

$$f_{code(i)}(x) = \begin{cases} 1/(2T_s) & x \in [-T_s, T_s] \\ 0 & otherwise \end{cases}. \quad (9)$$

where, $T_s$ is the maximum time hopping position with $T_s \leq T_f$.

For the fading amplitude $\alpha_{k,l}$, it follows a Nakagami-m distribution with parameters $(\Omega, m)$ according to

$$pdf(\alpha_{k,l}) = \frac{2}{\Gamma(m_{k,l})} (\frac{m_{k,l}}{\Omega_{k,l}})^{m_{k,l}} \alpha_{k,l}^{2m_{k,l}-1} \exp(-\frac{m_{k,l}}{\Omega_{k,l}} \alpha_{k,l}^2). \quad (10)$$

where, $\Gamma(\cdot)$ corresponds to the Gamma function, $m$ is the Nakagami m-factor which is modeled as a lognormally distribution random variable, $E[\alpha_{k,l}^2] = \Omega_{k,l}$.

The mean power of different rays is expressed by

$$E[\alpha_{k,l} \alpha_{k_1,l_1}] = \begin{cases} \dfrac{\Omega_l \exp(-\tau_{k,l}/\gamma_l)}{\gamma_l[(1-\beta)\lambda_1 + \beta\lambda_2 + 1]} & k = k_1 \text{ and } l = l_1 \\ 0 & k \neq k_1 \text{ or } l \neq l_1 \end{cases}. \quad (11)$$

where, $\Omega_l$ corresponds to the integrated energy of the $l$-th cluster, and $\gamma_l$ is the intra-cluster decay time constant. $\gamma_l$ is linearly depended on the arrival time of the cluster,

$$\gamma_l \propto k_\gamma T_l + \gamma_0 \quad (12)$$

and the mean energy of the $l$-th cluster is given by

$$10\log(\Omega_l) = 10\log(\exp(-T_l / \Gamma)) + M_{cluster} \quad (13)$$

III. PERFORMANCE ANALYSIS OF UWB BASED WSN

Considering a UWB system with time hopping multiple access and BPSK modulation, the transmitted signal can be expressed as:

$$s^{(n,i)}(t) = \sum_{j=0}^{N_s-1} d^{(n,i)} \sqrt{E_p} p(t - jT_f - C_{i,j}^{(n)} T_c). \quad (14)$$

where, $p(t)$ is a normalized pulse waveform, and the transmitted puse energy is $E_n$, $T_f$ is the frame duration, $C_{i,j}^{(n)} = \{1, 2, 3 \cdots, N_h\}$ is the time hopping sequence of the $j$-th frame of the $i$-th bit of the $n$-th user, $T_c$ is the time-hopping slot time, $d^{(n,i)} \in \{-1, 1\}$ represents the transmitted binary data sequence, and one data symbol is conveyed using $N_s$ pulses.

Thus, the received signal in time domain can be demonstrated as:

$$r(t) = \sum_{n=1}^{N_u} \sum_{i=1}^{N_l} s^{(n,i)}(t - \tau^{(n)} + iT_f) * h(t) + n(t). \quad (15)$$

where, $n(t)$ is the addictive white Gaussian noise and $\tau^{(n)}$ is the $n$-th user's reference delay relative to first user because of asynchronous transmission, and without loss of generalization, the first user's delay can be asssigned to be 0, i.e., $\tau^{(1)} = 0$. $N_I$ represents the number of interfering pulses from the previous periods, which can be defined as $N_I = \lceil \tau_{max}/T_f \rceil = \lceil \tau_{max} R_b N_s \rceil$, where, $R_b$ is the transmission data rate, and $\tau_{max}$ is the maximum multipath delay.

In the indoor office LOS environment, the mean ray interval $\Delta_\tau$ can be expressed using (16).

$$E[\Delta_\tau] = \int_0^{+\infty} \Delta_p \lambda_2 e^{-\lambda_2 \Delta_p} d\Delta_p = \frac{1}{\lambda_2}. \quad (16)$$

Since in indoor office LOS environment $\lambda_2 = 2.97(1/\text{ns})$, and then according to (16), the mean ray interval is approximately 0.34ns. However, in UWB based wireless sensor networks, the pulse duration is in the order of nanosecond, and hence, the transmitted pulse duration is bigger than the mean ray interval, which may result in the intra-symbol interference (IASI).

In the receiver end, it is assumed that the signal of the first ray from the first cluster of user 1 is to be received, and the template for demodulation is defined as:

$$v(t) = \sum_{j=0}^{N_s-1} p(t - jT_f - C_j^{(1)}T_c - T_1 - \tau_{1,1}). \quad (17)$$

When using correlation receiver, the output decision variable is given by

$$Z = \int_{iT_f}^{(i+1)T_f} r(t)v(t)dt = Z_u + Z_n + Z_{IASI} + Z_{ISI} + Z_{MUI}. \quad (18)$$

where, $Z_u$, $Z_n$, $Z_{IASI}$, $Z_{ISI}$, $Z_{MUI}$ account for the desired signal, additive white Gaussian noise, intra-symbol interference (IASI), inter-symbol interference (ISI) and multiuser interference (MUI), respectively. The energy for these components can be obtained using the following formulaitons:

$$E_b = E(Z_u)^2 = F(\omega_0)\Omega_0 N_s^2 \quad (19)$$

$$\sigma_n^2 = E(Z_n)^2 = F(\omega_0)N_s N_0/2 \quad (20)$$

$$\sigma_{IASI}^2 = F(\omega_0)N_s^2 \sum_{k=2}^{K} \int_0^{T_m} \Omega_0 \exp(-y/\gamma_l) f_p(y) R^2(y) dy. \quad (21)$$

$$\sigma_{ISI}^2 = E(Z_{ISI})^2 = F(\omega_0)N_s^2 \sum_{k=2}^{K} \int_{-T_m}^{T_m} \Omega_\Sigma e^{-y/\gamma_l} f_p(y) R^2(y) dy. \quad (22)$$

$$\sigma_{MUI}^2 = E(Z_{MUI})^2$$
$$= F(\omega_0)R_b N_s^2 N_u \sum_{k=2}^{K} \int_{-T_f/2}^{T_f/2} \int_{-z}^{T_m-z} \Omega_0 e^{-y/\gamma_l} f_p(y) R^2(y+z) dy dz \quad (23)$$
$$+ F(\omega_0)R_b N_s^2 N_u \sum_{k=2}^{K} \int_{-T_f/2}^{T_f/2} \int_{-z}^{T_m-z} \Omega_\Sigma e^{-y/\gamma_l} f_p(y) R^2(y+z) dy dz$$

where, $R(\cdot)$ represents the autocorrelation function of the transmitted pulse waveform, and $\Omega_0$ and $\Omega_\Sigma$ are given by (24) and (25), resepctively.

$$\Omega_0 = \frac{1}{\gamma_0[(1-\beta)\lambda_1 + \beta\lambda_2 + 1]}. \quad (24)$$

$$\Omega_\Sigma = \sum_{s=1}^{N_I N_s - 1} E[\Omega_s]$$
$$= \frac{1}{2T_s} \sum_{l=1}^{L} \sum_{s=1}^{N_I N_s - 1} \int_{-T_s}^{T_s} \int_0^{sT_f + \tau_{code}} \int_{sT_f + \tau_{code}}^{\tau_{max}} \Omega_0 \, e^{-T_l/\Gamma} \, e^{-(sT_f + \tau_{code} - T_l)/\gamma} \quad (25)$$
$$\times f_c(T_l)f_c(T_{l+1})dT_l dT_{l+1} d\tau_{code}.$$

where, $\Omega_s$ represents the energy from the $s$-th interfering pulse.

Finally, the signal to interference plus noise ratio (SINR) can be written as

$$\text{SINR} = \frac{E_b}{\sigma_n^2 + \sigma_{IASI}^2 + \sigma_{ISI}^2 + \sigma_{MUI}^2}. \quad (26)$$

And for BPSK modulated system, the bit error rate (BER) is given by:

$$BER = \frac{1}{2} erfc(\sqrt{\frac{\text{SINR}}{2}}). \quad (27)$$

IV. PERFORMANCE ANALYSIS

In IEEE 802.15.4a standard, two frequency band are allocated to UWB based WSN, i.e., [3244MHz, 4742MHz] and [5944MHz, 10234MHz]. In order to obain the pulse waveform complying with the spectrum requirement, the semi-definite programming (SDP) based pulse shaping method proposed in [9] is used. Gaussian pulse is used as the kernel function, and the number of the synthetic pulse is 19, and the time interval between two pulses are 0.04ns. Fig. 1 and Fig. 2 shows the obtained time domain and frequency domain pulse waveforms supported in the two frequency bands, respectively.

Using the obtained pulses shown in Fig. 1, simulation results and theoritical analysis of BER performance for UWB based WSN are givn in Fig. 3 using real line and dotted line, respectively. From Fig. 3, it is easy to find that the proposed system model and BER formulation can be used to evaluate system performance because the curve of theoritical analysis are very close to that of simulation results. The difference of these two curves are due to the reason that in simulations the sampling rate of pulse waveform is not enough high so that the distortion happens to pulse waveforms. When improve the sampling rate of pulse wavefomrs, the curve of simulations will get closer to the curve of theoritical analysis.

Moreover, from Fig. 3, the system performance will be affected by the transmitted pusle waveform, and in most situations, the BER performance of pulse waveform supported in [3244MHz, 4742MHz] is better than that in [5944MHz, 10234MHz] because that the pulse waveform supported in [3244MHz, 4742MHz] has little intra-symbol interference.

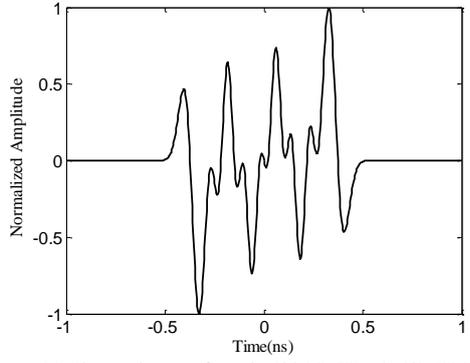

(a) The pulse waveform in [3244MHz, 4742MHz]

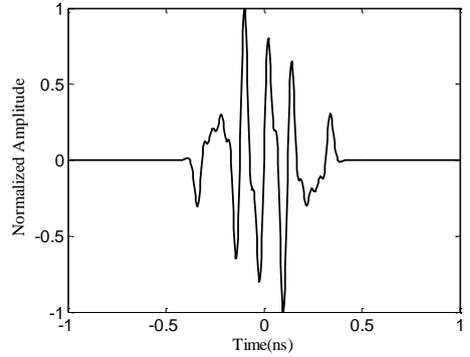

(b) The pulse waveform in [5944MHz, 10234MHz]

Figure 1.  Time domain pulse waveforms

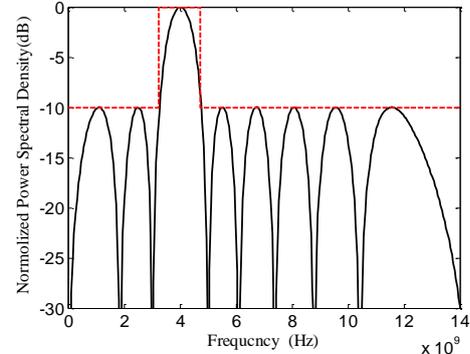

(a) The pulse waveform in [3244MHz, 4742MHz]

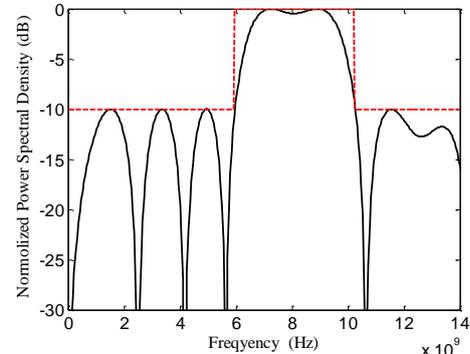

(b) The pulse waveform in [5944MHz, 10234MHz]

Figure 2.  Frequency domain pulse waveforms

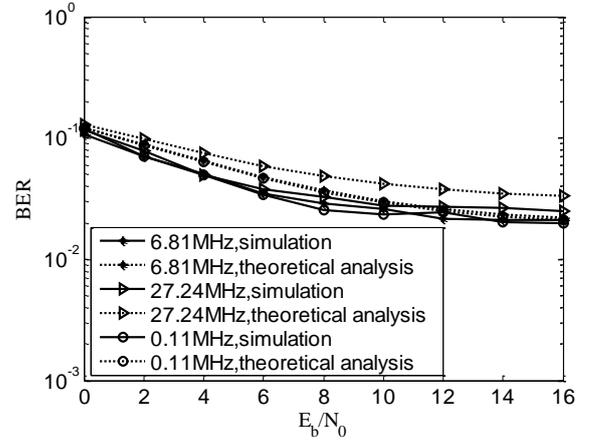

(a) The pulse waveform in [3244MHz, 4742MHz]

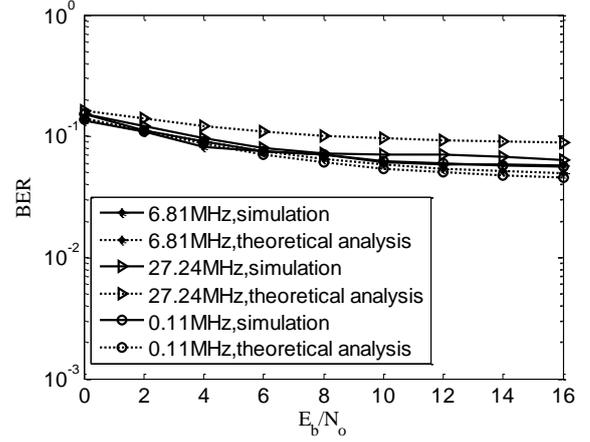

(b) The pulse waveform in [5944MHz, 10234MHz]

Figure 3.  Simulation results and theoritical analysis of BER performance

Furthermore, in Fig. 3, with the decrease of transmission data rate, the performance decreases slowly. When the transmission data rate are 6.81MHz and 0.11MHz, respecitively, the BER are nearly the same, and this is because that the intra-symbol interference is much higher than ISI. To evaluate the effect of the intra-symbol interference in UWB based WSN, under the assumption that the transmitted pulse energy is normalilzed to 1, Table 1 to Table 3 show the value of IASI, ISI and MUI in the transmission data rate of 27.24Mbps, 6.81Mbps, and 0.11Mbps, respectively.

TABLE I.  THE TRANSMISSION DATA RATE IS 27.24MBPS

|  | 1 user | 8 users | 16 users |
|---|---|---|---|
| **IASI** | *0.0288* | *0.0288* | *0.0288* |
| **ISI** | *0.0069* | *0.0069* | *0.0069* |
| **MUI** | *0* | *7.174\*10^{-24}* | *2.870\*10^{-23}* |

TABLE II.  THE TRANSMISSION DATA RATE IS 6.81MBPS

|  | 1 user | 8 users | 16 users |
|---|---|---|---|
| **IASI** | *0.0288* | *0.0288* | *0.0288* |
| **ISI** | *5.8119\*10^{-4}* | *5.8119\*10^{-4}* | *5.8119\*10^{-4}* |
| **MUI** | *0* | *5.713\*10^{-24}* | *2.285\*10^{-23}* |

TABLE III. THE TRANSMISSION DATA RATE IS 0.11MBPS

|       | 1 user | 8 users | 16 users |
|-------|--------|---------|----------|
| IASI  | 0.0288 | 0.0288  | 0.0288   |
| ISI   | 0      | 0       | 0        |
| MUI   | 0      | $5.610 \times 10^{-24}$ | $2.244 \times 10^{-23}$ |

From the above Tables, it is easy to find that with the decline of transmission data rate and user number, the ISI and MUI decreases. However, the intra-symbol interference (IASI), which became the main factor which degrade the system performance, remain the same in different enviroment due to the reason that the IASI reflect the effect of different multipaths in one bit.

Hence, in UWB based WSN, the IASI is the main reason for the decrease of system performance, and can be reduced by selecting pulse waveforms with better autocorrelation functions, which will be considered in future work.

## V. CONCLUSION

Based on the channel model in indoor office enviroment of IEEE 802.15.4a, a system model considering intra-symbol interference, inter-symbol interference, multiuser interference and addictive white Gaussian noise is proposed in this paper. Then, a SDP based pusle shaping method is used to obtain the pulse waveforms complying with the IEEE 802.15.4a spectrum requirement, and using the obtained pulses, the proposed system model and BER formulation is verified. Moreover, the parameters of transmission data rate and user number are also analyzed, and results show that in high SNR range, the intra-symbol interference is the main factor which degrade the system BER performance significantly even with low transmission data rate and small user number.

In future work, the effect of transmitted pulse waveform will be analyzed, and the pulse shaping methods will be throughly investigated and find the optimal pulse waveforms to minimize the effect of IASI.